\newcommand{\m}{\tau}
\documentstyle[12pt]{article}
\begin{document}
\begin{titlepage}

\centerline{\large \bf QCD SUM RULES CALCULATION}
\vspace{5mm}
\centerline{\large \bf OF THE SINGLET AXIAL CONSTANT}
\vspace{10mm}

\centerline{\large \bf A.V.Belitsky, O.V.Teryaev}
\vspace{10mm}

\centerline{\large \it Bogoliubov Laboratory of Theoretical Physics}
\vspace{3mm}
\centerline{\large \it Joint Institute for Nuclear Research}
\vspace{3mm}
\centerline{\large \it 141980, Dubna, Russia}
\vspace{20mm}

\centerline{\large \bf Abstract}
\large

We analyze the singlet axial form factor of the proton for small momentum
transferred in the framework of QCD sum rules using the interpolating
nucleon current which explicitly accounts for the gluonic degrees of
freedom. As the result we come to the quantitative description of
the singlet axial constant.
\end{titlepage}
%%%%%%%%%%%%%%%%%%%%%%%%%%%%%%%%%%%%%%%%%%%%%%%%%%%%%%%%%%%%%%%%%%%%%%%%%
\newpage
\large
\begin{flushleft}
{\bf1. Introduction.}
\end{flushleft}
\vspace{0.6cm}

The investigation of polarized deep inelastic scattering is one of
the most attractive field for theoretical consideration since it
provides an important insight into the structure of hadrons and
opens a large area of subtle dynamical phenomena associated with the
spin dependent case.
In the last several years there has been an increasing interest in the
deep inelastic structure function $g_1^p(x)$. It was provoked by the
EMC result on the scattering of the longitudinally polarized muon
beam on a longitudinally polarized hadron target. The unexpectedly
small asymmetry found by EMC has led to the so called "spin crisis in
the parton model" and has raised a number of questions of
understanding the dynamics of the proton spin on the parton level,
namely, how the nucleon spin is build up from the spins of
its constituents. An enormous flood of theoretical investigations
was generated in order to resolve the current "spin problem" \cite{efr94}.

The EMC measurement of the first moment of the polarized structure
function ${\Gamma}^p_1$ can be interpreted, via the Ellis-Jaffe
sum rule \cite{ell74}
\begin{eqnarray}
&&\!\!\!\!\!\!{\Gamma}_1^p({{\cal Q}^2})\!=\!\int_{0}^{1}dx g_1^p(x,{{\cal Q}^2})\nonumber\\
&&\!\!\!\!\!\!=\!\frac{1}{12} \Biggl\{\! \left(\!G_{A}^{(3)}(0)\!+\!\frac{1}{\sqrt{3}}G_{A}^{(8)}(0)\!\right)\! \left(\!1\!-\! \left(\frac{{\alpha}_{s}}{\pi}\right)\!-\!3.5833{\left(\frac{{\alpha}_{s}}{\pi}\right)}^2\!-\!20.2153{\left(\frac{{\alpha}_{s}}{\pi}\right)}^3\right)\nonumber\\
&&\;\;\qquad +\frac{4}{3}G_{A}^{(0)}(0,{{\cal Q}^2})\! \left(\!1\!-\!\frac{1}{3}\!\left(\frac{{\alpha}_{s}}{\pi}\right)\!-\!1.0959{\left(\frac{{\alpha}_{s}}{\pi}\right)}^2 \right)\!\Biggr\}, \label{elljaf}
\end{eqnarray}
as a first ever measurement of the singlet axial constant
$G_{A}^{(0)}(0)$ of the proton. The last one turns out to be
unexpectedly small in contradiction with the na\"\i ve
parton model where it is fairly close to unity. The EMC reported
the result for $G_{A}^{(0)}(0)$ which is compatible with zero. The
new experiments are performed to check their measurement of $g_1^p(x)$
and to measure an analogous neutron function $g_1^n(x)$. The recent
analysis \cite{exp95} of proton and deuteron data gives
$G_{A}^{(0)}(0)$ varying from $0.20\pm 0.11$ to $0.36\pm 0.05$, 
that is still far from unity.
So the problem reduces to the evaluation of $G_{A}^{(0)}(0)$
because the other two axial constants can be extracted reliably
from the data on deuteron (or $He^3$) 
target and on neutron and hyperon $\beta$-decays.
In this paper we calculate it in the framework of QCD sum rule approach
which till now seems to be the most powerful method for extraction of
information about the low energy properties of hadrons and
the closest one to the first principles of the theory.

In eq.(\ref{elljaf}) the functions $G_{A}^{(i)}(Q^2)$ are form
factors at zero momentum transferred in the proton matrix elements
of axial currents
\begin{eqnarray}
&&\!\!\!\!\!\!\!\!\!\!\!\!\!\!\!\!\langle N(p_2,{\lambda}_2)| j_{5\mu}^{(i)}(0) |N(p_1,{\lambda}_1) \rangle \nonumber\\
&&\qquad \qquad =\bar u_{N}^{({\lambda}_2)}(p_2) \left(G_{A}^{(i)}(Q^2){\gamma}_{\mu}{\gamma}_5-G_{P}^{(i)}(Q^2)q_{\mu}{\gamma}_5 \right) u_{N}^{({\lambda}_1)}(p_1), \label{formfac}
\end{eqnarray}
where $i$ is a $SU(3)_{f}$ index, $q=p_2-p_1$ and $Q^2=-q^2$.
There is an important
difference in the behaviour of induced pseudoscalar form factors
at small momentum $q$. Here, the singlet pseudoscalar form factor does
not acquire a Goldstone pole at $Q^2=0$, even in the chiral limit,
contrary to the matrix elements of the octet currents. It is known
that this limit, in which the masses of the three light quark
flavours are neglected, is not far away from the real world
of hadrons. In this limit, there exist eight massless pseudoscalar
mesons serving as Goldstone bosons. However, the ninth pseudoscalar,
the ${\eta}'$-meson, remains massive. In the following this property
will be used to extract a value of $G_{A}^{(0)}(0)$
from the sum rules.

It has been established \cite{efr94} that the first moment ${\Gamma}^p_1$
does not measure the contribution of the quark spins to the
proton one. This happens due to the anomalous nonconservation
of the singlet axial current. For this reason, we display the
r\^ole of this profound feature of the theory from the
very beginning exploiting the equation for the anomalous
divergence\footnote{\large Throughout the paper, we adopt the conventions
in Itzykson and Zuber \cite{itz80}.}:
\begin{equation}
{\partial}_{\mu}j_{5\mu}^{(0)}=2i\sum_{q}m_{q}\bar q {\gamma}_5q-\left(\! \frac{N_{f}{\alpha}_{s}}{4\pi}\! \right)G_{\mu \nu}^a\widetilde G_{\mu \nu}^a,
\end{equation}
where $N_{f}$ is a number of flavours (later $N_{f}=3$).
Taking the divergence of eq.(\ref{formfac}) for the singlet axial
current and making use of the last expression we come to the
relation which directly connects, in the chiral, limit the nonforward
matrix elements of the gluon operator to the effective form factor
$2m_{N}G_{eff}^{(0)}(Q^2)=2m_{N}G_{A}^{(0)}(Q^2)+Q^2G_{P}^{(0)}(Q^2)$
that is equal to the $2m_{N}G_{A}^{(0)}(0)$ at $Q^2=0$.

\newpage
\vspace{0.6cm}
\begin{flushleft}
{\bf 2. Effective axial form factor in QCD sum rules.}
\end{flushleft}
\vspace{0.6cm}

For a long time all calculations of the nucleon characteristics
use a particular three-quark current introduced by Ioffe \cite{iof81}.
When one makes an attempt to evaluate the matrix elements of
quark-gluon or gluon operators, one faces evident calculational
difficulties, moreover, the final sum rules are aggravated
by extra UV logarithms due to mixing of operators and, therefore, the
calculations are affected by noncontrollable uncertainties \cite{iof94}.

In field theory, the usual statement that the nucleon mainly consists
of three quarks means that the {\it 3 quarks $\rightarrow$ 3 quarks}
Green function (three quarks are in a state with nucleon quantum numbers)
has a pole at the mass of the nucleon, with the total angular momentum
$J=\frac{1}{2}$, with a significant residue. The fact that the nucleon
is not just a three quark state means that the nucleon pole also occurs,
albeit with smaller residue, in Green functions such as
{\it 3 quarks$+$g $\rightarrow$ 3 quarks$+$g}. For this reason, one is
forced to introduce a more sophisticated interpolating proton field which
explicitly contains the gluonic degrees of freedom:
\begin{equation}
{\eta}_G(x)={\epsilon}^{ijk}\left( u^i(x)C{\gamma}_{\mu} u^j(x)\right){\gamma}_5{\gamma}_{\mu}{\sigma}_{\alpha \beta}{\left( gG^a_{\alpha \beta}(x)t^ad(x)\right)}^k.
\end{equation}
The latter was investigated in ref. \cite{bra93} and checked in the
calculation of proton gluonic form factor normalized to the fraction
of nucleon momentum carried by gluons. Recently, making use of this
current the twist-3 and twist-4 corrections to Bjorken and Ellis-Jaffe
sum rule have been found \cite{ste95}. The advantages of this current are
straightforward: the calculations are drastically simplified, sum rules are
free from additional divergences that are not removed by the single
Borel transformation. At the same time, the applicability 
of non-dimensional regularization may be spoiled by the power UV
divergencies, appearing due to the high mass dimnension of this current.
For the same reason, sum rules may be affected by the vacuum condensates of 
higher dimensions, reducing their outcome to semiquantitative estimates. 
 
The usual technique of QCD sum rules is to extract the nucleon
matrix element of local operator from the appropriate three-point
correlation function. This correlator is the sum of different tensor
structures each characterized by the relevant invariant amplitude
$W^{(i)}(p_1^2,p_2^2,q^2)$.
\newpage
\begin{eqnarray}
&&\!\!\!\!\!\!\!\!W(p_1,p_2,q)\nonumber\\
&&=i^2 \int d^4xd^4ye^{ip_1x-ip_2y}\langle 0|T\left\{ {\eta}_G(x)   \left(\! \frac{N_{f}{\alpha}_{s}}{4\pi}\! \right)\!\!G_{\mu \nu}^a(0)\widetilde G_{\mu \nu}^a(0)\,  {\bar {\eta}}_G(y)\right\} |0\rangle\nonumber\\
&&\!\!\!\!\!\!\!\!={\sigma}_{\mu \nu}{\gamma}_5p_{1 \mu}p_{2 \nu}W^{(1)}(p_1^2,p_2^2,q^2)+{\not \! q}{\gamma}_5W^{(2)}(p_1^2,p_2^2,q^2)+q^2{\gamma}_5W^{(3)}(p_1^2,p_2^2,q^2).\nonumber\\  \label{expansion}
&&
\end{eqnarray}
In practical calculation it is advantageous to consider
$W^{(1)}(p_1^2,p_2^2,q^2)$ (hereafter referred to as $W$) because of its
lower dimensionality. Another
reason in favour of this choice is that it does not lead to the
fictitious kinematical singularities in $q^2$ as the last term
in eq.(\ref{expansion}) does. For this invariant amplitude
we can write the double dispersion representation
\begin{equation}
W(p_1^2,p_2^2,q^2)=
\frac{1}{{\pi}^2}\int_{0}^{\infty }\!\!\!\int_{0}^
{\infty }\!\!ds_{1}ds_{2}\frac{{\rho}(s_{1},s_{2},Q^2)}
{(s_{1}-p_1^2)(s_{2}-p_2^2)}+ \cdots,
\end{equation}
where the ellipses stand for the polynomials in $p_1^2$ and $p_2^2$
which die out after the double Borel transformation has been applied.
For the physical spectral density we have accepted the conventional
"resonance plus continuum" model:
\begin{eqnarray}
&&{\rho}(s_{1},s_{2},Q^2)={\pi}^2 m_{N}^{4}{\lambda}_{G}^{2}2m_{N}G_{eff}^{(0)}(Q^2)\delta (s_{1}-m_N^2)\delta (s_{2}-m_N^2)\nonumber\\
&&\qquad \qquad \qquad +{\rho}_{cont}(s_{1},s_{2},Q^2)(1-\theta ({\sigma}_0-s_1)\theta ({\sigma}_0-s_2)).
\end{eqnarray}
The function in front of the double-pole term is a combination of form factors
we are interested in up to certain overlap ${\lambda}_{G}$ between the state
created from the vacuum by ${\eta}_G$ and the nucleon state
\begin{equation}
\langle 0| {\eta}_G(0) |N(p,{\lambda}) \rangle =m_N^2{\lambda}_G u_{N}^{({\lambda})}(p).
\end{equation}
So, our aim is the evaluation of the correlation function (\ref{expansion})
in QCD. In the case when all the momenta $(-p_1^2)\sim (-p_2^2)\sim Q^2$
are sufficiently large (of an order of $1 {GeV}^2$), the leading contribution
comes from the domain where all distances are small. Thus, the standard
machinery of short distance expansion are applicable, allowing one to express
the final result in terms of quark and gluon condensates. The problem
is modified drastically if the squared momentum transferred becomes
small ($Q^2\ll (-p_i^2)$) because the relevant $t$-channel distances
can be large. In this case the OPE has a twofold structure \cite{bal82}.
Terms of the first type arise from the SD(I)-region when all intervals
$x^2\sim y^2\sim (x-y)^2$ are small. Another contribution comes from
SD(II)-region (bilocal power correction) which originates from the distances
$x^2\sim y^2\gg (x-y)^2$.
The necessity for the bilocal power corrections can be traced from the
fact that the ordinary QCD Feynman diagrams contributing to the form factor
at moderately large $Q^2$ in the limit of small $Q^2$ possess
logarithmic non-analyticities $(Q^2)^nlnQ^2$, which signals that large
distances come into play \cite{ama78}. Therefore, we have to subtract
such a {\it perturbative} behaviour from the corresponding graphs
and add the "exact" correlators which account for the nonperturbative
effects and thus possess the correct analytical properties as $Q^2$
goes to zero. So the OPE in the case when the momentum
transferred can be arbitrary small has a modified form \cite{bal82,nes84}
\begin{eqnarray}
&&\!\!\!\!\!\!\!\!\!W(p_1^2,p_2^2,q^2)\nonumber\\
&&\!\!\!\!\!\!\!\!\!=\sum_{d}C_{SD(I)}^{(d)}(p_1^2,p_2^2,q^2)
\langle {\cal O}_d \rangle
+\sum_{i} \int d^4x e^{ipx}C_{SD(II)}^{(i)}(x^2) W_i^{BL}(x,q),
\end{eqnarray}
where, as was mentioned above, the coefficients
$C_{SD(I)}^{(d)}(p_1^2,p_2^2,q^2)$ are
regular in the limit $Q^2\rightarrow 0$. The second term determines the
large $t$-distance contribution. Here $W_i^{BL}$ are the two-point
correlators
\begin{equation}
W_i^{BL}(x,q)=\int d^4y e^{iqx}\langle 0|T\left\{ {\cal G}(y) {\cal O}_i(x,0) \right\} |0\rangle
\end{equation}
of operator in question
${\cal G}(y)\!=\!\left(\! \frac{N_{f}{\alpha}_{s}}{4\pi}\! \right)\!G_{\mu \nu}^a
(y)\widetilde G_{\mu \nu}^a(y)$
and some nonlocal string operator with definite
twist (not dimension) \cite{bal89,rus93}
that arises from the OPE of the $T$-product of nucleon
currents:
\begin{equation}
T\left\{ {\eta}_G(x) {\bar {\eta}}_G(0)\right\}
=\sum_{i} C_{SD(II)}^{(i)}(x^2){\cal O}_i(x,0)
\end{equation}
The bilocal power corrections cannot be directly calculated
in perturbation theory but we can write down the dispersion relation
for them
\begin{equation}
W_i^{BL}(x,q)=\frac{1}{\pi}\int_{0}^{\infty}ds\frac{{\rho}_i(s,(xq),x^2)}{s-q^2}, \label{biloc}
\end{equation}
assuming the standard spectral density model with continuum to start at
some threshold $s_0$ and finding in some way its parameters. We always
do this constructing auxiliary sum rules. There is no need in
additional subtractions in eq.(\ref{biloc}) because one always deals
with the difference between the "exact" bilocal and its perturbative
part; so due to the coincidence of their UV behaviours the subtraction terms
cancel in this difference.

To simplify the calculation of the local power corrections,
it is convenient to use fixed-point gauge for the background
gluon field $(x-x_0)_{\mu}B^a_{\mu}(x)=0$. We chose the
fixed point in the vertex of the gluon operator $x_0=0$.
The quark and gluon propagators in this gauge up to the order $O(G)$
looks like \cite{nov84}
\begin{eqnarray}
&&\!\!\!\!\!\!\!\! D^{ab}_{\mu \nu}(x,y)\nonumber\\
&&\!\!\!\!\!\!\!\! =-i\frac{\Gamma (\frac{d}{2}\!-\!1)}
{4{\pi}^{\frac{d}{2}}}
\frac{g_{\mu \nu}{\delta}^{ab}}{[-{\Delta}^2]^{\frac{d}{2}-\!1}}
+2i G^{ab}_{\mu \nu} \frac{\Gamma (\frac{d}{2}\!
-\!2)}{16{\pi}^{\frac{d}{2}}}\frac{1}{[-{\Delta}^2]^{\frac{d}{2}
-\!2}}+iG^{ab}_{\rho \sigma} \frac{\Gamma
(\frac{d}{2}\!-\!1)}{8{\pi}^{\frac{d}{2}}}
\frac{g_{\mu \nu}y_{\rho}x_{\sigma}}{[-{\Delta}^2]^{\frac{d}{2}-\!1}},
\nonumber\\
&&\!\!\!\!\!\!\!\! S(x,y)\nonumber\\
&&\!\!\!\!\!\!\!\! =\frac{\Gamma (\frac{d}{2})}{2{\pi}^{\frac{d}{2}}}
\frac{{\not \!\! {\Delta}}}{[-{\Delta}^2]^{\frac{d}{2}}}
-\widetilde G_{\mu \nu} {\gamma}_{\nu}{\gamma}_5
\frac{\Gamma (\frac{d}{2}\!-\!1)}{8{\pi}^{\frac{d}{2}}}
\frac{{\Delta}_{\mu}}{[-{\Delta}^2]^{\frac{d}{2}-\!1}}
+iG_{\mu \nu}y_{\mu}x_{\nu} \frac{\Gamma (\frac{d}{2})}
{4{\pi}^{\frac{d}{2}}} \frac{{\not \!\!
{\Delta}}}{[-{\Delta}^2]^{\frac{d}{2}}},
\end{eqnarray}
where $\Delta =x-y$, $G^{ab}_{\mu \nu}=gf^{acb}G^{c}_{\mu \nu}$
for the gluon propagator and
$G_{\mu \nu}=gt^aG^a_{\mu \nu}$ for the quark one, the generators
are normalized by $Sp(t^a t^b)=\frac{1}{2}{\delta}^{ab}$.
For the noncollinear quark condensate we use the following expansion
in terms of local vacuum expectation values \cite{gro95}:
\begin{eqnarray}
&&\!\!\!\!\!\!\!\!\!\!\!\!\langle
\bar{\psi}^i_{\alpha}(y){\psi}^i_{\beta}(x) \rangle=
\frac{1}{4}\langle \bar{\psi}{\psi} \rangle I_{\beta \alpha}
\nonumber\\
&&
+ \frac{1}{4^3}m_{0}^2 \langle \bar{\psi}{\psi}\rangle \left[ (x-y)^2
-i\frac{2}{3}{\sigma}_{\mu \nu}x_{\mu}y_{\nu} \right]_{\beta \alpha}+...
\end{eqnarray}
The ellipses stand for the higher dimension vacuum condensates.

Calculating the diagrams depicted in the first row of fig. 1
we come to the Borel sum rule for effective axial form factor
at moderate values of the momentum transferred with the
Borel parameters ${\m}_1$ and ${\m}_2$:

\begin{eqnarray}
&&\!\!\!\!\!\!{{m_{N}^{4}{\lambda}_{G}^{2}}
\over {{\m}_{1}{\m}_{2}}}2m_{N}G_{eff}^{(0)}(Q^2)
e^{-\frac{m_{N}^2}{{\m}_{1}}-\frac{m_{N}^2}{{\m}_{2}}}\nonumber\\
&&\!\!\!\!\!\!=-\frac{1}{{\pi}^2}\int_{0}^{\infty}\!\!\!
\int_{0}^{\infty}\!{{ds_{1}ds_{2}} \over {{\m}_{1}{\m}_{2}}}
(1-\theta ({\sigma}_0-s_1)\theta ({\sigma}_0-s_2))
{\rho}_{cont}(s_{1},s_{2},Q^2)e^{-\frac{s_{1}}{{\m}_{1}}
-\frac{s_{2}}{{\m}_{2}}}\nonumber\\
&&\!\!\!\!\!\!\!\!\!+\frac{N_f}{{\pi}^2}
{\left(\frac{{\alpha}_{s}}{\pi}\right)}^2
\langle\bar u u\rangle
\Bigl\{
\frac{1}{3}Q^2
\frac{({\m}_{1}{\m}_{2})^2}{({\m}_{1}\!\!+\!\!{\m}_{2})^3}
J_{12}\nonumber\\
&&\!\!\!\!\!\!\!\!\!+
\frac{1}{18}
m_{0}^2
\biggl[ 4\frac{({\m}_{1}{\m}_{2})^2}
{({\m}_{1}\!\!+\!\!{\m}_{2})^3}
J_{12}-Q^2\frac{({\m}_{1}{\m}_{2})}{({\m}_{1}\!\!+\!\!{\m}_{2})^2}
( 4J_{11}-J_{02})
\biggr]\nonumber\\
&&\!\!\!\!\!\!\!\!\!+
\frac{1}{144}
m_{0}^2
\biggl[
4\frac{({\m}_{1}{\m}_{2})^2}{({\m}_{1}\!\!+\!\!{\m}_{2})^3} J_{12}
+Q^2\frac{({\m}_{1}{\m}_{2})}{({\m}_{1}\!\!+\!\!{\m}_{2})^2}
J_{02}
\biggr]\nonumber\\
&&\!\!\!\!\!\!\!\!\!-
\frac{7}{32}
m_{0}^2
\frac{({\m}_{1}{\m}_{2})^2}{({\m}_{1}\!\!
+\!\!{\m}_{2})^3} J_{12}
%\nonumber\\
%&&\!\!\!\!\!\!\!\!\!
-\frac{1}{8}
m_{0}^2
\biggl[ 2
\frac{({\m}_{1}{\m}_{2})^2}{({\m}_{1}\!\!+\!\!{\m}_{2})^3} J_{12}
-Q^2\frac{({\m}_{1}{\m}_{2})}{({\m}_{1}\!\!+\!\!{\m}_{2})^2}
J_{02}
\biggr]\Bigr\}
\label{modff}
\end{eqnarray}

where the continuum double spectral density is
\begin{equation}
{\rho}_{cont}(s_{1},s_{2},Q^2)=\sum_{i=1}^{5}{\rho}_{(i)}(s_{1},s_{2},Q^2),
\end{equation}
and each term in a sum is found from the corresponding diagram in fig.1
(appropriate formulae are given in appendix B):
\begin{eqnarray}
&&\!\!\!\!\!\!\!\!\!\!\!\!\!\!\!\!\!\!\!\!\!\!\!\!\!\!\!\!\!\!\!\!\!
\!\!\!\!\!\!\!\!\!\!\!\!\!\!\!\!{\rho}_{(1)}(s_{1},s_{2},Q^2)
=\frac{N_{f}}{72}{\left(\frac{{\alpha}_{s}}{\pi}\right)}^2
\langle \bar u u\rangle Q^4 {\left(1
-\frac{\sigma}{R^{\frac{1}{2}}}\right)}^2\left(2
+\frac{\sigma}{R^{\frac{1}{2}}}\right),
\end{eqnarray}
\begin{eqnarray}
&&\!\!\!\!\!\!\!\!\!\!\!\!\!\!\!\!\!\!\!\!\!\!\!
\!\!\!\!\!{\rho}_{(2)}(s_{1},s_{2},Q^2)
=\frac{N_{f}}{18}{\left(\frac{{\alpha}_{s}}{\pi}\right)}^2\!
m_{0}^2\langle \bar u u\rangle Q^2\biggl\{\frac{1}{6}{\left(1\!
-\!\frac{\sigma}{R^{\frac{1}{2}}}\right)}^2\!\!\left(2\!
+\!\frac{\sigma}{R^{\frac{1}{2}}}\right)\!\nonumber\\
&&\qquad \qquad \qquad \qquad \qquad \qquad -\frac{1}{2}\left(1\!
-\!\frac{\sigma}{R^{\frac{1}{2}}}\right)-5Q^2
\frac{s_{1}s_{2}}{R^{\frac{3}{2}}}\!\biggr\},\nonumber\\
&&
\end{eqnarray}
\begin{eqnarray}
&&\!\!\!\!\!\!\!{\rho}_{(3)}(s_{1},s_{2},Q^2)\nonumber\\
&&\!\!\!\!\!\!\!=\frac{N_{f}}{144}{\left(
\frac{{\alpha}_{s}}{\pi}\right)}^2\! m_{0}^2
\langle \bar u u\rangle Q^2\left\{\!\frac{1}{6}{\left(1\!
-\!\frac{\sigma}{R^{\frac{1}{2}}}\right)}^2\!\!\left(2\!
+\!\frac{\sigma}{R^{\frac{1}{2}}}\right)\!-\!\frac{1}{2}\left(1\!
-\!\frac{\sigma}{R^{\frac{1}{2}}}\right)\!
-\!Q^2 \frac{s_{1}s_{2}}{R^{\frac{3}{2}}}\!\right\},\nonumber\\
&&
\end{eqnarray}
\begin{eqnarray}
&&\!\!\!\!\!\!\!\!\!\!\!\!\!\!\!\!\!\!\!\!\!\!\!\!\!
\!\!\!\!\!\!\!\!{\rho}_{(4)}(s_{1},s_{2},Q^2)
=-\frac{7N_{f}}{768}{\left(\frac{{\alpha}_{s}}{\pi}\right)}^2
m_{0}^2\langle \bar u u\rangle Q^2 {\left(1
-\frac{\sigma}{R^{\frac{1}{2}}}\right)}^2\left(2
+\frac{\sigma}{R^{\frac{1}{2}}}\right),
\end{eqnarray}
\begin{eqnarray}
&&\!\!\!\!\!\!\!\!{\rho}_{(5)}(s_{1},s_{2},Q^2)\nonumber\\
&&\!\!\!\!\!\!\!\!=\!-\frac{N_{f}}{8}{\left(
\frac{{\alpha}_{s}}{\pi}\right)}^2\! m_{0}^2
\langle \bar u u\rangle Q^2\left\{\!\frac{1}{12}{\left(1\!
-\!\frac{\sigma}{R^{\frac{1}{2}}}\right)}^2\!\!\left(2\!
+\!\frac{\sigma}{R^{\frac{1}{2}}}\right)\!+\!\frac{1}{2}\left(1\!
-\!\frac{\sigma}{R^{\frac{1}{2}}}\right)\!
+\!Q^2 \frac{s_{1}s_{2}}{R^{\frac{3}{2}}}\!\right\}\!,\nonumber\\
&&
\end{eqnarray}
and
\begin{equation}
\sigma=s_{1}+s_{2}+Q^2,
\hspace{1cm}R(s_{1},s_{2},Q^2)={\sigma}^2-4s_{1}s_{2}.
\end{equation}

The functions $J_{nm}$ are originated from the diagrams in the first
row in fig.1 and are given by the following expression:
\begin{equation}
J_{nm}({\m}_i,Q^2)
=\int_{0}^{1}dx{\bar x}^{n-1}x^m exp
\left\{ -\frac{x}{\bar x}\frac{Q^2}{({\m}_{1}\!+\!{\m}_{2})} \right\}
\end{equation}

We state that contrary to the refs.\cite{ste95} where the sum rules
with the same interpolating nucleon field were
dominated by the contribution from the highest dimension operators,
our sum rule is not affected by them: the coefficient functions
that are determined to the leading accuracy by tree and one-loop
diagrams vanish identically. Therefore, we do not meet the problem of
breakdown of OPE for the correlator in question. The absence
of higher condensates contribution unsuppressed by a number of loops is
directly connected with chiral structure of the interpolating nucleon
field and the tensor structure chosen for investigation
in the three-point correlation function.

As was mentioned previously one could not put $Q^2=0$ in the
eq. (\ref{modff}) because though finite it is including contributions
non-analytic in this point. This is typical example of the mass
singularities. Therefore, following the method outlined in the middle
of this section one should subtract the perturbative behaviour
from corresponding graphs (the diagrammatic representation for them
is shown in the second row of fig. 1, while the explicit expressions
are written in the appendix A) and add the terms with correct singular
structure in $Q^2$. It is clear that the singularity should be located at the
threshold of the first prominent resonance in the corresponding channel.

\vspace{0.6cm}
\begin{flushleft}
{\bf 3. Bilocal corrections.}
\end{flushleft}
\vspace{0.6cm}

The simplest bilocal correction (first picture on fig. 2) is given
by the convolution of
the coefficient function involving the quark condensate with
the two-point correlation function of operator in question and
some point-splitted gluon operator coming from the OPE of nucleon
fields

\begin{eqnarray}
&&\!\!\!\!\!\!\!\!\!{W^{BL}}_{\lambda \kappa} (x,q)\nonumber\\
&&\!\!\!\!\!\!\!\!\!=\!i\left(\!
\frac{N_{f}{\alpha}_{s}}{4\pi}\!\right)^2\!\!\!\int\!\!
d^4y e^{iqy}\langle 0|T\!\left\{ \!G_{\mu \nu}^a(y)
\widetilde G_{\mu \nu}^a(y)\!\!\left(\!G_{\mu \lambda}^a(x)
\widetilde G_{\mu \kappa}^a(0)\!-\!G_{\mu \kappa}^a(x)
\widetilde G_{\mu \lambda}^a(0)\!\right)\!\right\}\!\!|0
\rangle \nonumber\\
&&
\end{eqnarray}

As was mentioned earlier we cannot calculate it in perturbation theory
but we can reconstruct it from the information about its large-$Q^2$
behaviour. To this end we write down the dispersion relation for
it of the type represented by eq. (\ref{biloc}) and use the standard
"resonance plus continuum" spectral density model, with ${\eta}'$-meson.
It is likely to be the only prominent {\it singlet} pseudoscalar both
in quark and gluon channels.

\begin{equation}
\rho (s)=\pi m_{{\eta}'}^2 f_{{\eta}'}
[if^{(1)}_{{\eta}'}{\phi}^{(1)}_{{\eta}'}(xq)\!
+\!f^{(2)}_{{\eta}'}(xq){\phi}^{(2)}_{{\eta}'}(xq)]
\delta (s-m_{{\eta}'}^2)
+\theta (s-s_0)\rho^{PT}(s).
\end{equation}
Here we have parametrized the matrix elements of some gluon operators
between the vacuum and ${\eta}'$-meson state as follows:
\begin{eqnarray}
&&\!\!\!\!\!\!\! \langle 0|\frac{N_{f}{\alpha}_{s}}{4\pi}
G_{\mu \nu}^a(0)\widetilde G_{\mu \nu}^a(0)|{\eta}'(q)\rangle
=m_{{\eta}'}^2 f_{{\eta}'} \nonumber\\
&&\!\!\!\!\!\!\! \langle{\eta}'(q) |
\frac{N_{f}{\alpha}_{s}}{4\pi}\!\!\left(G_{\mu \lambda}^a(x)
\widetilde G_{\mu \kappa}^a(0)\!-\!G_{\mu \kappa}^a(x)
\widetilde G_{\mu \lambda}^a(0) \right)\!\!|0\rangle\nonumber\\
&&=(x_{\kappa}q_{\lambda}\!-\!x_{\lambda}q_{\kappa})
[if^{(1)}_{{\eta}'}{\phi}^{(1)}_{{\eta}'}(xq)\!
+\!f^{(2)}_{{\eta}'}(xq){\phi}^{(2)}_{{\eta}'}(xq)].
\label{param}
\end{eqnarray}

In the last line the wave functions ${\phi}^{(i)}_{{\eta}'}$ can be
related in the standard way to the usual ones (${\varphi}^{(i)}$),
describing the light-cone momentum fraction distribution of gluon
inside meson.
\begin{equation}
{\phi}^{(i)}(xq)
=\int_{0}^{1}d{\alpha} e^{i{\alpha}(xq)}{\varphi}^{(i)}(\alpha).
\end{equation}
In eq. (\ref{param}) we have kept only the leading twist wave functions
which reproduce the leading nonanalyticity in the corresponding
contribution of the local $\langle\bar u u\rangle$ power correction.
As will be shown below, we account for other contribution using simple
recipe which results from our consideration (originally it was proposed
in Ref. \cite{bel93}).
We can find the unknown overlaps $f^{(i)}_{{\eta}'}$ constructing the
auxiliary sum rules. It turns out, that due to the antisymmetrical
tensor structure involved the contribution of ordinary local power
corrections with gluon condensates, unsuppressed by a loop factor,
vanish identically in the theoretical part of the sum rules.
For this reason we account for nonperturbative
effects introducing the concept of nonlocal gluon condensate \cite{gro82}
which corresponds to infinite series of local ones. It can be appropriately
decomposed into two tensor structures multiplied by corresponding form
factors \cite{gro95,sim88}:
\begin{eqnarray}
&&\!\!\!\!\!\!\!\!\!\!\!\!\!\!\!\!\!\!\langle 0|
G_{\mu \rho}^a(x)\widetilde E^{ab}(x,0)
G_{\nu \sigma}^b(0)|0\rangle\nonumber\\
&&\!\!\!\!\!\!\!\!\!\!\!\!\!\!\!\!\!\!
=\frac{\langle G^2\rangle}{12}\biggl\{(g_{\mu \nu}g_{\rho \sigma}
-g_{\mu \sigma}g_{\nu \rho})\left(D_{NA}(x^2)+D_{A}(x^2)\right)\nonumber\\
&&\qquad +(g_{\mu \nu}x_{\rho}x_{\sigma}
+g_{\rho \sigma}x_{\mu}x_{\nu}-g_{\mu \sigma}x_{\nu}x_{\rho}
-g_{\nu \rho}x_{\mu}x_{\sigma})\frac{dD_{A}(x^2)}{dx^2}
\biggr\}. \label{nonlocal}
\end{eqnarray}
This form explicitly separates out the term proportional to $D_{NA}(x^2)$
which violate the abelian Bianchi identity, while the second term
satisfies it. It was shown that linear confinement occurs when $D_{NA}(x^2)$
is present in (\ref{nonlocal}) while the second term does not contribute
to the string tension \cite{sim88}.

In the calculation of $f^{(1)}_{{\eta}'}$ and $f^{(2)}_{{\eta}'}$
constants only abelian form factor contribute. We present it in the
form of $\alpha$-representation \cite{mik93}:
\begin{equation}
D_{A}(x^2)
=\int_{0}^{\infty}d\alpha f_{G}^{A}(\alpha,
{\lambda}_{A}^2)e^{\alpha \frac{x^2}{4}}.\label{form}
\end{equation}
and use a $\delta$-shaped ansatz for the distribution function
$f_{G}^{A}(\alpha,{\lambda}_{A}^2)$:
\begin{equation}
f_{G}^{A}(\alpha,{\lambda}_{A}^2)=\delta(\alpha
-{\lambda}_{A}^2),\label{distribution}
\end{equation}
where $1/{\lambda}_{A}$ is an abelian correlation length of the vacuum
fluctuations, it can be expressed in terms of vacuum condensates
${\lambda}_{A}^2=\frac{8}{9}g^2{
\langle \bar u u\rangle}^2/{\langle G^2\rangle} \approx 0.03 GeV^2$
at $1GeV^2$ \cite{nik83}. One comment concerning eq. (\ref{form}) is that
in deriving a QCD sum rule one can always perform a Wick rotation
$x_0\rightarrow ix_0$ and treat all the coordinates as Euclidean,
$x^2<0$.

Proceeding in the standard way we obtain the following sum rule:
\begin{eqnarray}
&&\!\!\!\!\!\!\!\!\!\!\!\!\!\!\!\!\!\!m_{{\eta}'}^2
f_{{\eta}'}f^{(1)}_{{\eta}'}e^{-\frac{m_{{\eta}'}^2}{M^{2}}}\nonumber\\
&&\!\!\!\!\!\!\!\!\!\!\!\!\!\!\!\!\!\!=M^{2}\frac{N_{f}{\alpha}_{s}}{4\pi}
\left( \frac{1}{3{\pi}^2}\frac{N_{f}{\alpha}_{s}}{4\pi}M^{4}E_2
\left(\frac{s_0}{M^{2}}\right)
+\frac{N_{f}}{3}\langle \frac{{\alpha}_{s}}{\pi}G^2 \rangle
\frac{{\lambda}_{A}^2}{M^{2}}
\left(1-\frac{{\lambda}_{A}^2}{M^{2}} \right) \right),
\label{srule}
\end{eqnarray}
where
\begin{equation}
E_2(x)=1-(1+x+\frac{x^2}{2})e^{-x},
\end{equation}

We stress that due to the fact that one of the gluon currents has
nonzero Lorentz spin leads to the absence of direct instantons to
the polarization operator of interest \cite{nov81}.
This property may be considered as a counterpart of the absence
of the topological ghost pole in the formfactor of interest \cite{efr94},
while this pole does appear in the formfactor of the conserved 
quark-gluon current, related to the low-energy nucleon structure.    

We take now the limit $M^2 \rightarrow \infty$ and obtain the local duality
relation:
\begin{eqnarray}
m_{{\eta}'}^2 f_{{\eta}'}f^{(1)}_{{\eta}'}
=\frac{N_{f}{\alpha}_{s}}{4\pi}
\left( \frac{1}{3{\pi}^2}\frac{N_{f}{\alpha}_{s}}{4\pi}\frac{s_0^3}{6}
+\frac{N_{f}}{3}\langle \frac{{\alpha}_{s}}{\pi}G^2 \rangle
{\lambda}_{A}^2 \right),
\end{eqnarray}
The value of continuum threshold is found from the requirement
of the most stable sum rule (\ref{srule}). Straightforward
analysis gives us the value  $s_0=2.5{GeV}^2$ which coincides
with the one obtained in ref.\cite{iof92}.

Keeping the contribution due to the nonlocal gluon condensate
would exceed the accuracy we are pretending to because,
as it was mentioned above, we do not
calculate the corresponding term in the OPE with local power
corrections which are obviously small. The leading non-zero
contribution coming from non-local condensate was required to 
analyze the stability of the sum rule and to determine the continuum
threshold. 
However, the relative numerical value of the non-local condensate 
contribution is small (like that of the dropped local term), 
and we neglect it in what follows. 

For the overlap factor $f^{(2)}_{{\eta}'}$
an analogous relation taken to the same accuracy is:
\begin{eqnarray}
m_{{\eta}'}^2 f_{{\eta}'}f^{(2)}_{{\eta}'}
=-\frac{1}{420{\pi}^2}\left(\frac{N_{f}{\alpha}_{s}}{4\pi}\right)^2
\frac{s_0^3}{6}
\end{eqnarray}

The net form for the additional term for the axial form factor at
small momentum transferred looks like
\begin{eqnarray}
&&m_N^5\lambda_G^2\delta G_{eff}^{(0)}(Q^2)
e^{-\frac{m_N^2}{M^2}}\nonumber\\
&&=\frac{2}{3}\langle \bar u u\rangle M^4e^{\frac{Q^2}{4M^2}}
\biggl\{
\frac{N_f}{(4\pi)^2}
{\left(\frac{{\alpha}_{s}}{\pi}\right)}^2
\left[
Q^4\ln \left( \frac{s_0+Q^2}{Q^2}\right) -s_0Q^2+\frac{s_0^2}{2}
\right]\nonumber\\
&&\qquad\qquad\qquad-\frac{4}{N_f}
\frac{m_{{\eta}'}^2 f_{{\eta}'}}{Q^2+m_{{\eta}'}^2}
[f^{(1)}_{{\eta}'}{\varphi}^{(1)}_{{\eta}'}(1/2)\!
+\!f^{(2)}_{{\eta}'}{\dot\varphi}^{(2)}_{{\eta}'}(1/2)]
\biggr\}\nonumber\\
&&=\frac{2}{3}\langle \bar u u\rangle M^4 e^{\frac{Q^2}{4M^2}}
\frac{N_f}{(4\pi)^2}{\left(\frac{{\alpha}_{s}}{\pi}\right)}^2\nonumber\\
&&\qquad\qquad\qquad\left\{
Q^4\ln \left( \frac{s_0+Q^2}{Q^2}\right) -s_0Q^2
+\frac{s_0^2}{2}-\frac{s_0^3/3}{Q^2+m_{{\eta}'}^2}
\right\}\label{add}
\end{eqnarray}

In the last line we have substitute the local duality relations for
residue factors and take the asymptotical form for the wave functions
\begin{eqnarray}
&&{\varphi}^{(1)}(\alpha)=30{\alpha}^2{\bar {\alpha}}^2.\nonumber\\
&&{\varphi}^{(2)}(\alpha)
=420(\alpha\!-\!\bar {\alpha}){\alpha}^2{\bar {\alpha}}^2
\end{eqnarray}

Note that in eq. (\ref{add}) we have put $\m_1=\m_2=2M^2$ in order
not to introduce the asymmetry between the initial and final states
and to make contact with Borel parameter of the two-point nucleon
sum rules. We can observe that at large $Q^2$ the bilocals vanish
faster than the term it is correcting $\sim Q^2$. Note, that
although the local power correction may vanish for $Q^2\to 0$
the modified version of nonanalyticities $Q^{2n}\ln Q^2$ alive
in this limit.

As can be easily seen, the
leading nonanalyticity in the $\langle\bar u u\rangle $-term
in the correction found and in the expression for form factor cancels. 
It is replaced by the combination $s_0+Q^2$ which
is "safe" in the limit $Q^2\rightarrow 0$. In the same way we may
correct the other nonanalyticities.
In large $Q^2$ limit where the original OPE must be valid the bilocal
corrections must be absent. As we have seen, the residues of the
physical spectrum can be found from the requirement that the bilocal
power corrections should vanish faster at large $Q^2$ than the
contribution they are correcting. Then, in general the correction term
is given by the following equation (we omit all unnecessary constants)

\begin{equation}
W_{BL}^{res+cont}-W_{BL}^{PT}=
\frac{s_0^n/n}{Q^2+m_{{\eta}'}^2}-\int_{0}^{s_0}\frac{ds s^{n-1}}{s+Q^2}.
\end{equation}

Using this simple recipe one can easily modify all perturbative
non-analyticities of the effective axial form factor in the small
momentum transferred limit. After all of them have been corrected
properly we can take the limit $Q^2\to 0$ and obtain the sum rule
for singlet axial constant directly. As was noted in Ref.
\cite{bra93}, one should not try to fix the parameter of continuum
threshold $\sigma_0$ from the sum rule with a new current. 
Therefore we take
the continuum threshold usual for the sum rules involving the
nucleon and consider it in the local duality limit. Combining all
contribution we come to the following equation
\begin{eqnarray}
&&m_N^5{\bar\lambda}_G^2G_A^{(0)}
=N_fa
\left(
\frac{\alpha_s}{\pi}
\right)^2
\biggl\{
\frac{7}{2^33}m_0^2\frac{\sigma_0^3}{6}
+\left[\frac{1}{3}R_3
+ \frac{31}{2^5}m_0^2R_2\right]\frac{\sigma_0^2}{2}\nonumber\\
&&-\left[\frac{1}{2^23}R_4 + \frac{751}{2^73^2}m_0^2R_3\right]\sigma_0
+\left[\frac{1}{2^53^2}R_5 + \frac{1479}{2^{10}3^3}m_0^2R_4\right]
\biggr\}
\end{eqnarray}
where
\begin{equation}
R_n=
\left[
\frac{s_0^n}{nm_{\eta'}^2}-\frac{s_0^{n-1}}{n-1}
\right]
\end{equation}
and its origin was clarified above.
The sum rule imply that we model the continuum by effective spectral
density that includes all ones which are nonzero for $s>0$.

We use the standard ITEP values of condensates rescaled to the
normalization point ${\mu}^2\sim m^2_N\sim 1{GeV}^2$ with the appropriate
anomalous dimensions: $a=-(2\pi)^2\langle\bar u u\rangle =0.67 GeV^3$,
$m_{0}^2=\langle \bar u g(\sigma G) u\rangle
/\langle \bar u u\rangle =0.65GeV^2$; also, we use the overlap value 
${\bar\lambda}_G^2=2(2\pi)^4\lambda_G^2=0.3 GeV^6$
and continuum threshold ${\sigma}_0=2.5GeV^2$, providing the 
better accuracy of the calculation of partonic densities \cite{bra3}. 
The value of the strong
coupling constant at $1GeV^2$ is taken to be $\alpha_s=0.37$
that corresponds to $\Lambda =150 MeV$. We obtain the following
numerical value of the singlet axial constant
\begin{equation}
G_{A}^{(0)}(0)=0.2.
\end{equation}
Varying the parameters in the reasonable range will result in the
variation of the quantity within the $50\%$. The main uncertainties
come from the errors in estimation of the $t$-channel continuum
threshold $s_0$ and the overlap $\lambda_G$ of the nucleon state with
that created by the new current.

Due to the anomalous non-conservation of the singlet axial current
the singlet axial constant is not a renormalization group invariant.
Therefore in order to compare our prediction with the experimentally
mensurable quantity, we have to evolve it from QCD sum rule scale
${\mu}^2\sim 1 {GeV}^2$
up to the one of EMC-SMC experiment which is ${\cal Q}^2=10 {GeV}^2$
exploiting the one-loop solution of RG equation:
\begin{equation}
G_{A}^{(0)}(0,{\cal Q}^2)=G_{A}^{(0)}(0,{\mu}^2)
exp\left\{ \frac{{\gamma}_2}{4{\pi}{\beta}_0}
[{\alpha}_s({\cal Q}^2)-{\alpha}_s({\mu}^2)] \right\},
\end{equation}
where the anomalous dimension ${\gamma}_2=16N_f$ and as usual
${\beta}_0=11-\frac{2}{3}N_f$. However, the sensitivity to
the QCD radiative corrections is poor until very large $Q^2$
is attained and account for them would exceed the accuracy of
our estimate. Nevertheless, the value obtained are in reasonable
agreement with the new world average value for the singlet axial
constant.

\vspace{0.6cm}
\begin{flushleft}
{\bf 4. Summary.}
\end{flushleft}
\vspace{0.6cm}

In summary, we have calculated the singlet axial constant in the QCD
sum rule framework for the form factor type problem at small momentum
transferred and find the value in good correspondence with experimental
one. We should mention that in our letter \cite{bel96} $G_A^{(0)}$ was
somewhat overestimated because the next-to-leading twist bilocal
power corrections were not accounted for and the continuum was
not properly subtracted. 

In ref.\cite{iof92} the pioneering attempt was 
undertaken to evaluate $G_{A}^{(0)}(0)$
by QCD sum rules in a way similar to the calculation of the octet axial
constant \cite{bel85}. Due to the presence of the gluon anomaly in the
induced vacuum condensates the problem differs significantly from the one
for the $G_{A}^{(8)}(0)$. This feature was incorporated in the calculation
but nevertheless the authors did not come to the reasonable quantitative
prediction of the singlet axial constant. It was conjectured that
the OPE breaks down for the singlet axial current in the
{\it {axial-nucleon-nucleon}} vertex. At the same, we did not observe
any evidence of the divergence of the OPE in the correlator under
investigation: the contribution of the highest dimension vacuum
condensates unsuppressed by a number of loops is absent.
From the other side, the small values of the power corrections result
in the good accuracy of the local duality approach and in the strong 
dependence of our result on the continuum threshold. 

A possible line of development would be to estimate twist-three
gluon contribution into the moments of the transverse spin structure
function $g_2$ \cite{bel94} and the $x$-dependence of the 
twist-two polarized gluon distribution
in the nucleon. However, the latter would require an elaboration of
the new procedure for separation of the large and
small distances in the effective four-point correlator.

\vspace{0.6cm}
\begin{flushleft}
{\bf Acknowledgments.}
\end{flushleft}
\vspace{0.6cm}

We would like to thank  A.E.Dorokhov, A.V.Efremov, B.L.Ioffe,
N.I. Ko\-che\-lev, L.Man\-kie\-wicz, S.V.Mikhailov, A.V.Radyushkin,
R.Ruskov and A.Sch\"a\-fer for useful conversations.
This work was supported by the International
Science Foundation under grant RFE300 and the Russian Foundation for
Fundamental Investigation under grant $N$ 93-02-3811.

\vspace{0.6cm}
\begin{flushleft}
{\bf Appendix A.}
\end{flushleft}
\vspace{0.6cm}

In this appendix we present some useful integrals needed in the
additional factorization of large and small distances in the ordinary
Feynman diagrams. Keeping the lowest twists contributions into the
perturbative parts of the bilocal correlators we find:

\begin{eqnarray}
&&\!\!\!\!\!\!\!\!\!\int\!\!\frac{d^dk}{(2\pi)^d}
\frac{(x(k+\tilde q))^n}{[k^2-l]^r}
=\frac{i(-1)^r}{(4\pi)^{\frac{d}{2}}}\frac{1}{\Gamma(r)}
\biggl\{\!(x\tilde q)^n\frac{\Gamma(r\!-\!\frac{d}{2})}{l^{r
-\frac{d}{2}}}\nonumber\\
&&\!\!\!\!\!\!\!\!\!+\frac{1}{2}C_n^2[-x^2](x\tilde q)^{n-2}
\frac{\Gamma(r\!-\!\frac{d}{2}\!-\!1)}{l^{r-\frac{d}{2}-1}}\!
+\!\frac{3}{4}C_n^4[-x^2]^2(x\tilde q)^{n-4}\frac{\Gamma(r\!
-\!\frac{d}{2}\!-\!2)}{l^{r-\frac{d}{2}-2}}\!\biggr\}\!+\!O(x^6),\nonumber\\
&&
\end{eqnarray}

\begin{eqnarray}
&&\!\!\!\!\!\!\!\!\int\!\!\frac{d^dk}{(2\pi)^d}\frac{(x(k+\tilde q))^n
k_{\lambda}}{[k^2-l]^r}=\frac{i(-1)^{r-1}}{(4\pi)^{\frac{d}{2}}}
\frac{x_{\lambda}}{2\Gamma(r)}\biggl\{\!C_n^1(x\tilde q)^{n-1}
\frac{\Gamma(r\!-\!\frac{d}{2}\!-\!1)}{l^{r-\frac{d}{2}-1}}\nonumber\\
&&\!\!\!\!\!\!\!\!+\frac{3}{2}C_n^3[-x^2](x\tilde q)^{n-3}
\frac{\Gamma(r\!-\!\frac{d}{2}\!-\!2)}{l^{r-\frac{d}{2}-2}}\!
+\!\frac{15}{4}C_n^5[-x^2]^2(x\tilde q)^{n-5}\frac{\Gamma(r\!
-\!\frac{d}{2}\!-\!3)}{l^{r-\frac{d}{2}-3}}\!\biggr\}\!+\!O(x^6),\nonumber\\
&&
\end{eqnarray}

\begin{eqnarray}
&&\!\!\!\!\!\!\!\!\int\!\!\frac{d^dk}{(2\pi)^d}
\frac{(x(k+\tilde q))^n(kq)k_{\lambda}}{[k^2-l]^r}
=\frac{i(-1)^{r-2}}{(4\pi)^{\frac{d}{2}}}
\frac{x_{\lambda}}{2\Gamma(r)}\biggl\{\!C_n^2(xq)(x\tilde q)^{n-2}
\frac{\Gamma(r\!-\!\frac{d}{2}\!-\!2)}{l^{r-\frac{d}{2}-2}}\nonumber\\
&&\!\!\!\!\!\!\!\!+3C_n^4[-x^2](xq)(x\tilde q)^{n-4}
\frac{\Gamma(r\!-\!\frac{d}{2}\!-\!3)}{l^{r-\frac{d}{2}-3}}
+\frac{45}{4}C_n^6[-x^2]^2(xq)(x\tilde q)^{n-6}
\frac{\Gamma(r\!-\!\frac{d}{2}\!-\!4)}{l^{r-\frac{d}{2}-4}}\!
\biggr\}\nonumber\\
&&\!\!\!\!\!\!\!\!+\frac{i(-1)^{r-1}}{(4\pi)^{\frac{d}{2}}}
\frac{q_{\lambda}}{2\Gamma(r)}\biggl\{\!(x\tilde q)^n
\frac{\Gamma(r\!-\!\frac{d}{2}\!-\!1)}{l^{r-\frac{d}{2}-1}}\nonumber\\
&&\!\!\!\!\!\!\!\!+\frac{1}{2}C_n^2[-x^2](x\tilde q)^{n-2}
\frac{\Gamma(r\!-\!\frac{d}{2}\!-\!2)}{l^{r-\frac{d}{2}-2}}
+\frac{3}{4}C_n^4[-x^2]^2(x\tilde q)^{n-4}\frac{\Gamma(r\!
-\!\frac{d}{2}\!-\!3)}{l^{r-\frac{d}{2}-3}}\!\biggr\}\!+\!O(x^6),\nonumber\\
&&
\end{eqnarray}
where $\tilde q=\alpha q$, and $C_n^m=\frac{n!}{m!(n-m)!}$ are binomial
coefficients.

Using these results we obtain for the factorized diagrams in the
second row in fig.1:

\begin{eqnarray}
\!\!\!\!\!\!\!\!\overline{W}^{(1)}=\frac{1}{{\pi}^2}
\frac{N_{f}}{3}{\left(\frac{{\alpha}_{s}}{\pi}\right)}^2
\langle \bar u u\rangle \frac{({\m}_{1}{\m}_{2})^2}{({\m}_{1}\!\!
+\!\!{\m}_{2})^2}\left(L_4+L_6+\frac{1}{6}L_8\right)
e^{\frac{Q^2}{({\m}_{1}\!+\!{\m}_{2})}},
\end{eqnarray}

\begin{eqnarray}
&&\!\!\!\!\!\!\!\!\!\!\!\!\!\!\!\!\overline{W}^{(2)}
=\frac{1}{{\pi}^2}\frac{N_{f}}{18}
{\left(\frac{{\alpha}_{s}}{\pi}\right)}^2
m_0^2\langle \bar u u\rangle \frac{({\m}_{1}{\m}_{2})}{({\m}_{1}\!\!
+\!\!{\m}_{2})^3}\biggl([4({\m}_{1}{\m}_{2})
-({\m}_{1}\!\!+\!\!{\m}_{2})^2]L_2\nonumber\\
&&\!\!\!\!\!\!\!\!\!\!\!\!\!\!\!\!\quad\qquad+[4({\m}_{1}{\m}_{2})-6({\m}_{1}\!\!
+\!\!{\m}_{2})^2]L_4+\frac{1}{6}[4({\m}_{1}{\m}_{2})
-15({\m}_{1}\!\!+\!\!{\m}_{2})^2]L_6 \biggr)
e^{\frac{Q^2}{({\m}_{1}\!+\!{\m}_{2})}},
\end{eqnarray}

\begin{eqnarray}
&&\!\!\!\!\!\!\!\!\!\!\!\!\!\!\!\!\overline{W}^{(3)}
=\frac{1}{{\pi}^2}
\frac{N_{f}}{144}{\left(\frac{{\alpha}_{s}}{\pi}\right)}^2
m_0^2\langle \bar u u\rangle \frac{({\m}_{1}{\m}_{2})}{({\m}_{1}\!\!
+\!\!{\m}_{2})^3}\biggl([4({\m}_{1}{\m}_{2})-({\m}_{1}\!\!
+\!\!{\m}_{2})^2]L_2\nonumber\\
&&\!\!\!\!\!\!\!\!\!\!\!\!\!\!\!\!\quad\qquad+[4({\m}_{1}{\m}_{2})-2({\m}_{1}\!\!
+\!\!{\m}_{2})^2]L_4+\frac{1}{6}[4({\m}_{1}{\m}_{2})
-3({\m}_{1}\!\!+\!\!{\m}_{2})^2]L_6 \biggr)
e^{\frac{Q^2}{({\m}_{1}\!+\!{\m}_{2})}},
\end{eqnarray}

\begin{eqnarray}
\!\!\!\!\!\!\!\!\overline{W}^{(4)}=-\frac{1}{{\pi}^2}\frac{7N_{f}}{32}{\left(
\frac{{\alpha}_{s}}{\pi}\right)}^2
m_0^2\langle \bar u u\rangle \frac{({\m}_{1}{\m}_{2})^2}{({\m}_{1}\!\!
+\!\!{\m}_{2})^3}\left(L_2+L_4
+\frac{1}{6}L_6\right)e^{\frac{Q^2}{({\m}_{1}\!+\!{\m}_{2})}},
\end{eqnarray}

\begin{eqnarray}
&&\!\!\!\!\!\!\!\!\!\!\!\!\!\!\!\!\overline{W}^{(5)}
=-\frac{1}{{\pi}^2}\frac{N_{f}}{8}{\left(\frac{{\alpha}_{s}}
{\pi}\right)}^2 m_0^2\langle \bar u u\rangle
\frac{({\m}_{1}{\m}_{2})}{({\m}_{1}\!\!
+\!\!{\m}_{2})^3}\biggl([2({\m}_{1}{\m}_{2})
+({\m}_{1}\!\!+\!\!{\m}_{2})^2]L_2\nonumber\\
&&\!\!\!\!\!\!\!\!\!\!\!\!\!\!\!\!\quad\qquad+[2({\m}_{1}{\m}_{2})+2({\m}_{1}\!\!
+\!\!{\m}_{2})^2]L_4+\frac{1}{6}[2({\m}_{1}{\m}_{2})
+3({\m}_{1}\!\!+\!\!{\m}_{2})^2]L_6 \biggr)
e^{\frac{Q^2}{({\m}_{1}\!+\!{\m}_{2})}},
\end{eqnarray}
and
\begin{eqnarray}
L_n={\left(\frac{Q^2}{{\m}_{1}\!\!+\!\!{\m}_{2}}\right)}^{n/2}
\left[ln\Bigl(\frac{Q^2}{{\mu}^2}\frac{{\m}_{1}{\m}_{2}}{({\m}_{1}\!\!
+\!\!{\m}_{2})^2}\Bigr)-S_1(n/2-1)\right],
\end{eqnarray}
with
\begin{eqnarray}
S_1(\alpha)=\psi(1+\alpha)+{\gamma}_{E}.
\end{eqnarray}
where we have used the $\overline {MS}$-scheme.

As can be easily seen the logarithmic terms in eqs. $(A4-A8)$ reproduce
the leading non-analiticities in the expressions for ordinary diagrams
contributing to the form factor at the moderately large $Q^2$.
Therefore this perturbative "long-distance" behaviour cancels exactly
in the difference of the diagrams in the first and second rows in fig.1.

\vspace{0.6cm}
\begin{flushleft}
{\bf Appendix B.}
\end{flushleft}
\vspace{0.6cm}

Here we present the formulae which enable to represent the integrals
appeared in the calculations in the form of double spectral representation.

\begin{eqnarray}
&&\!\!\!\!\!\!\!\!\frac{({\m}_1{\m}_2)^2}{({\m}_1\!+\!{\m}_2)^2}J_{11}
=\int\limits_{0}^{\infty}ds_1ds_2 e^{-\frac{s_1}{{\m}_1}
-\frac{s_2}{{\m}_2}}\left[ Q^2\frac{s_1s_2}{R^{\frac{3}{2}}} \right],\\
&&\!\!\!\!\!\!\!\!\frac{({\m}_1{\m}_2)^2}{({\m}_1\!+\!{\m}_2)^2}J_{02}
=\int\limits_{0}^{\infty}ds_1ds_2 e^{-\frac{s_1}{{\m}_1}
-\frac{s_2}{{\m}_2}}\left[ -\frac{1}{2}\left(1-\frac{\sigma}{R^{\frac{1}{2}}}
\right)- Q^2\frac{s_1s_2}{R^{\frac{3}{2}}} \right],\\
&&\!\!\!\!\!\!\!\!\frac{({\m}_1{\m}_2)^3}{({\m}_1\!+\!{\m}_2)^3}J_{12}
=\int\limits_{0}^{\infty}ds_1ds_2 e^{-\frac{s_1}{{\m}_1}
-\frac{s_2}{{\m}_2}}\left[ \frac{1}{24}Q^2{
\left( 1-\frac{\sigma}{R^{\frac{1}{2}}}\right)}^2\left( 2
+\frac{\sigma}{R^{\frac{1}{2}}} \right)\right].
\end{eqnarray}

\newpage
\pagestyle{empty}
\begin{flushleft}
{\bf Figure captions.}
\end{flushleft}
\vspace{1cm}
\begin{flushleft}
{\bf Fig.1.} Contribution to the effective axial form factor
in the QCD sum rules approach.
The difference of the first and second rows defines the SD(I)-regime.
\end{flushleft}
\vspace{0.5cm}
\begin{flushleft}
{\bf Fig.2.}
Generic form of the bilocal power corrections entering the OPE
with different coefficient functions.
\end{flushleft}

\newpage
\pagestyle{empty}

\unitlength=2.10pt
\special{em:linewidth 0.4pt}
\linethickness{0.4pt}



\begin{thebibliography}{99}
\bibitem{efr94}
For a review, see Anselmino M., Efremov A.V., Leader E. //
Phys. Rept. 1995. V. 261. P. 1.\\
Ioffe B.L. Preprint ITEP-61, 1994.
\bibitem{ell74}
Ellis J, Jaffe R.L. // Phys. Rev. 1974. V. D9. P. 1444;
1974.V. D10. P. 1669.\\
Kodaira J. // Nucl. Phys. 1980. V. B165. P. 129.\\
Larin S.A., Vermaseren J.A.M. // Phys. Lett. 1991. V. 259B. P. 213.\\
Larin S.A. // Phys. Lett. 1994. V. 334B. P. 192.
\bibitem{exp95}
Adams D. et al. // Phys. Lett. 1995. V. B357. P. 284.\\
Abe K. et al. // Phys. Lett. 1995. V. B364. P.61.
\bibitem{itz80}
Itzykson C., Zuber J. Quantum Field Theory. McGraw-Hill, 1980.
\bibitem{iof81}
Ioffe B.L. // Nucl. Phys. 1981. V. B188. P. 317; 1981. V. B191. P. 591.
\bibitem{iof94}
Ioffe B.L. Preprint BUTP-94/25, 1994.
\bibitem{bra93}
Braun V.M., G\'ornicki P., Mankiewicz L., Sch\"afer A. // Phys. Lett. 1993.
V. 302B. P. 291.
\bibitem{ste95}
Stein E., G\'ornicki P., Mankiewicz L.,
Sch\"afer A., Greiner W. // Phys. Lett. 1995. V. 343B. P. 369.\\
Stein E., G\'ornicki P., Mankiewicz L.,
Sch\"afer A. // Phys. Lett. 1995. V. 353B. P. 107.
\bibitem{bal82}
Balitsky I.I. // Phys. Lett. 1982. V. 114B. P. 53.
\bibitem{ama78}
Amati D., Petronzio R., Veneziano G. // Nucl. Phys. 1978. V. 140B. P. 54.\\
Libby S.B., Sterman G. // Phys. Rev. 1978. V. D18. P. 3252.\\
Ellis R.K. et al. // Nucl. Phys. 1979. V. B152. P. 285.
\bibitem{nes84}
Balitsky I.I., Yung A.V. // Phys. Lett. 1983. V. 129B. P. 328.\\
Ioffe B.L., Smilga A.V. // Nucl. Phys. 1984, V. B232. P. 109.\\
Nesterenko V.A., Radyushkin A.V // JETP Lett. 1984. V. 39. P. 707.\\
Balitsky I.I., Kolesnichenko A.V., Yung A.V. // Phys. Lett. 1985. V. 157B. P. 309.
\bibitem{bal89}
Balitsky I.I., Braun V.M., Kolesnichenko A.V. //
Nucl. Phys. 1989. V. B312. P. 509.
\bibitem{rus93}
Radyushkin A.V., Ruskov R. // Phys. Atom. Nucl. 1993. V. 56. P. 630;
1995. V. 58. P.1440.
\bibitem{nov84}
Novikov V.A., Shifman M.A., Vainshtein A.I., Zakharov V.I. //
Fortschr. Phys. 1984. V. 32. P. 585.
\bibitem{gro95}
Grozin A.G. // Int. Jour. Mod. Phys. 1995 V. A10. P. 3497.
\bibitem{bel93}
Belyaev V.M., Kogan Ya.I. //  Preprint ITEP-29, 1984;
Int. Jour. Mod. Phys. 1993 V. A8. P. 153.
\bibitem{gro82}
Gromes D. // Phys. Lett. 1982. V. 115B. P. 482.
\bibitem{sim88}
Dosh H.G., Simonov Yu.A. // Phys. Lett. 1988. V. 205B  P. 338.\\
Simonov Yu.A. // Nucl. Phys. 1988. V. B307. P. 512; 1989. V. B324. P. 67.
\bibitem{mik93}
Mikhailov S.V., Radyushkin A.V // Phys. Rev. 1991. V. D45. P.1754.\\
Mikhailov S.V. // Phys. Atom. Nucl. 1993. V. 56. P. 650.
\bibitem{nik83}
Nikolaev S.N., Radyushkin A.V. // Nucl. Phys. 1983. V. B213. P. 285.
\bibitem{nov81}
Novikov V.A., Shifman M.A., Vainshtein A.I., Zakharov V.I. //
Nucl. Phys. 1981. V. B191. P. 301.
\bibitem{iof92}
Ioffe B.L., Khodzamiryan A.Yu. // Sov. J. Nucl. Phys. 1992. V. 55. P. 1701.
\bibitem{bra3}  Braun V., Gornicki P., Mankiewicz L.//
Phys. Rev. 1995. V. D51. P. 6036. 
\bibitem{bel96}
Belitsky A.V., Teryaev O.V. // Phys. Lett. 1996. V. 366B. P. 345.
\bibitem{bel85}
Belyaev V.M., Ioffe B.L., Kogan Ya.I. // Phys. Lett. 1985. V. 151B. P. 290.
\bibitem{bel94}
Belitsky A.V., Efremov A.V., Teryaev O.V. // Phys. Atom. Nucl. 1995.
V. 58. P.1333.
\end{thebibliography}
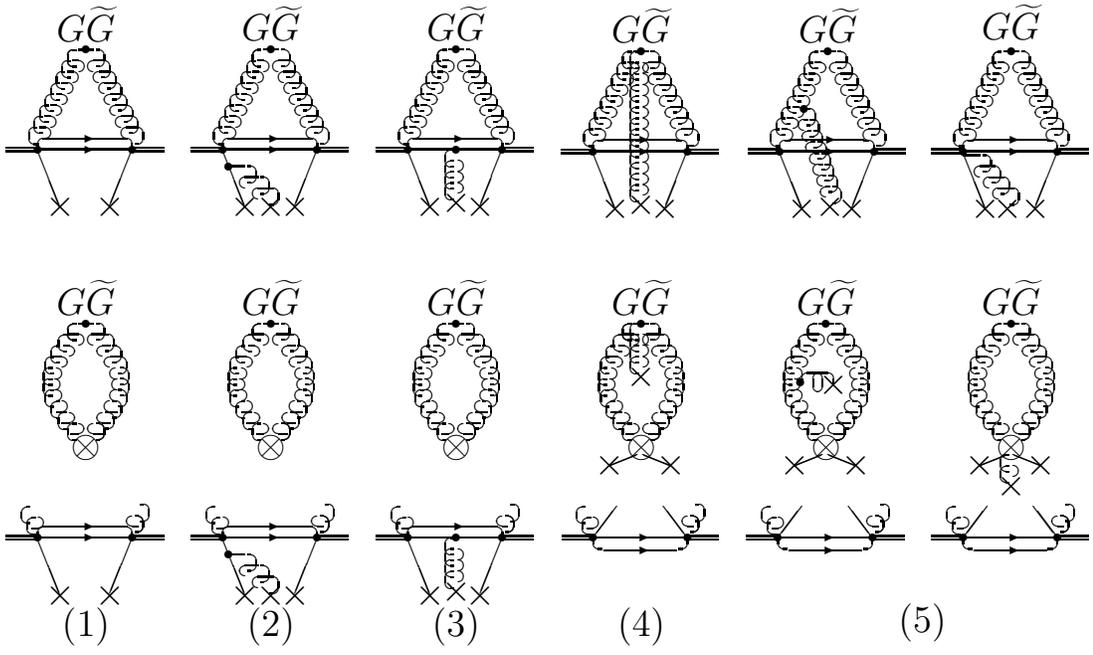
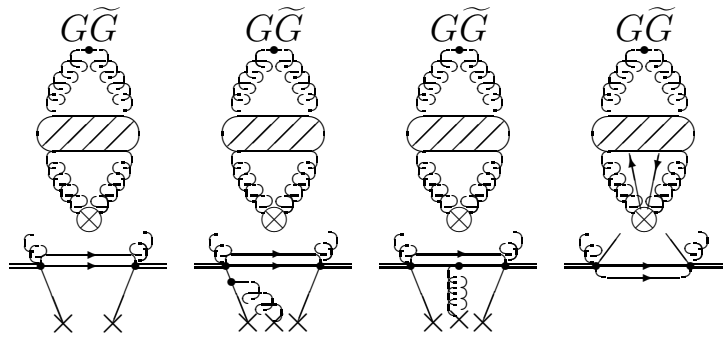
\end{document}